\documentclass[12pt]{llncs}

\newcommand{\mybar}[1]{\overline{#1}}

\newcommand{\eventra}[1]{\ensuremath{\langle #1 \rangle}}

\def\smallromani{\renewcommand{\theenumi}{\roman{enumi}}
\renewcommand{\labelenumi}{(\theenumi)}}

\newcommand{\Proof}{\NI
                    {\bf Proof.}\ }

\usepackage{stmaryrd}
\usepackage{sgame}
\usepackage{amsmath}
\usepackage{handbookbib}

\usepackage{color}
\usepackage{graphics}

\newcommand{\oldbfe}[1]{\begin{bfseries}\emph{#1}\end{bfseries}}

\newcommand{\ES}{\mbox{$\emptyset$}}

\newcommand{\myra}{\mbox{$\:\rightarrow\:$}}

\newcommand{\lra}{\mbox{$\:\leftrightarrow\:$}}

\newcommand{\A}{\mbox{$\ \wedge\ $}}

\newcommand{\sse}{\mbox{$\:\subseteq\:$}}

\newcommand{\fa}{\mbox{$\forall$}}
\newcommand{\te}{\mbox{$\exists$}}

\newcommand{\LL}{\mbox{$\ldots$}}

\newcommand{\newMS}[1]{\mbox{$[\![{#1}]\!]$}}

\newcommand{\NI}{\noindent}
\newcommand{\HB}{\hfill{$\Box$}}
\newcommand{\VV}{\vspace{5 mm}}

\newcommand{\restr}[1]{\! \mid \! {#1}}

\newcommand{\szkew}[1]{\relax \setbox0=\hbox{\kern -24pt $\displaystyle#1$\kern 0pt }%
\box0}
{\catcode`\@=11 \global\let\ifjusthvtest@=\iffalse}

\newcounter{oldmycaption}

\usepackage{amsfonts}
\usepackage{latexsym}
\usepackage{alltt}

\usepackage{proof}

\usepackage{rotating}

\title{Public Announcements in Strategic Games with Arbitrary Strategy Sets}

 \author{Krzysztof R. Apt \inst{1,2}
\and Jonathan A. Zvesper \inst{3}
 }
 \institute{CWI, Science Park 123, 1098 XG Amsterdam, The Netherlands
\and University of Amsterdam
\and Oxford University Computing Laboratory, 
Parks Road, Oxford OX1 3QD, UK
}

\date{}
\begin{document}

\date{}
\maketitle

\begin{abstract}
  In \cite{vB07} the concept of a public announcement is used to study
  the effect of the iterated elimination of strictly dominated
  strategies.  We offer a simple generalisation of this approach to
  cover arbitrary strategic games and many optimality notions. We
  distinguish between announcements of optimality and announcements of
  rationality.
\end{abstract}


\section{Introduction}


One of the objectives of the theory of strategic games is to predict the choices
players will make. This is usually done by studying iterated elimination of strategies
that one assumes players will not choose. In this paper we investigate to what
extent this elimination process can be explained using the concept of public announcements.

Public announcements were introduced into epistemic logic in
\cite{Pla89}.  In \cite{vB07} they were used to describe the process
of iterated elimination of strictly dominated strategies in a finite
game as a series of public announcements that transform an epistemic
model of the players.  In this paper we pursue this idea and obtain a
generalization of van Benthem's results to arbitrary strategic games
and to various different optimality operators that the players might
employ to select their strategies.

The paper is organized as follows.  In the next section we discuss the
notions we shall rely on.  Next, in Section \ref{sec:setup} we
introduce optimality properties that we discuss in the paper. We also
explain the crucial distinction between 'local' and 'global'
properties.  Globality means that each subgame obtained by iterated
elimination of strategies is analyzed \emph{in the context} of the
given initial game, while locality means that the context is the
current subgame.

In Section \ref{sec:opt}, we consider public announcements of
`optimality'.  We partially motivate van Benthem's use of what we call
`standard' models for games.  In Section \ref{sec:rat}, we consider
public announcements of `rationality' which are based on players knowledge.
This analysis shows the
relevance of the concept of global optimality properties.  

Unlike either of the two cited papers,
we do not examine any of the interesting logical questions of axiomatisation or definability.




\section{Preliminaries}
\label{sec:prelim}

In this section we introduce notions central to this note,
and recall some basic results concerning them.

\subsection{Operators}

Consider a fixed complete lattice $(D, \sse)$ with the largest element $\top$.
In what follows we use ordinals and denote them by $\alpha, \beta, \gamma$.
Given a, possibly transfinite, sequence $(G_{\alpha})_{\alpha < \gamma}$ of
elements of $D$ we denote their 
meet by $\bigcap_{\alpha < \gamma} G_{\alpha}$.

\begin{definition}
Let $T$ be an operator on $(D, \sse)$, i.e., $T: D \myra D$.

\begin{itemize}


\item We call $T$ \oldbfe{contracting} if for all $G$, $T(G) \sse G$.
\item We say that an element $G$ is a \oldbfe{fixpoint} of $T$ if $T(G) = G$.
\item We define by 
transfinite induction a sequence of elements $T^{\alpha}$ of $D$, where $\alpha$ is an ordinal, as follows:

\begin{itemize}

  \item $T^{0} := \top$,

  \item $T^{\alpha+1} := T(T^{\alpha})$,

  \item for limit ordinals $\beta$, $T^{\beta} := \bigcap_{\alpha < \beta} T^{\alpha}$.
  \end{itemize}

\item We call the least $\alpha$ such that $T^{\alpha+1} = T^{\alpha}$ the \oldbfe{closure ordinal} of $T$
and denote it by $\alpha_T$.  We call then $T^{\alpha_T}$ the \oldbfe{outcome of} (iterating) $T$ and write it alternatively as $T^{\infty}$.
\HB
\end{itemize}
\end{definition}

So an outcome is a fixpoint reached by a transfinite iteration that
starts with the largest element.  In general, the outcome of an
operator does not need to exist but we have the following obvious
observation.
%
%

\begin{note} \label{note:con}
Every contracting operator $T$ on $(D, \sse)$ has an outcome, i.e., 
$T^{\infty}$ is well-defined.
\HB
\end{note}

\subsection{Strategic games}

Given $n$ players ($n > 1$) by a \oldbfe{strategic game} (in short, a
\oldbfe{game}) we mean a sequence
$
(S_1, \LL, S_n, p_1, \LL, p_n),
$
where for all $i \in \{1, \LL, n\}$,
$S_i$ is the non-empty set of \oldbfe{strategies} available to player $i$,
and
$p_i$ is the \oldbfe{payoff function} for the  player $i$, so
$
p_i : S_1 \times \LL \times S_n \myra \cal{R},
$
where $\cal{R}$ is the set of real numbers.

We denote the strategies of player $i$ by $s_i$, possibly with some
superscripts.  Given $s \in S_1 \times \LL \times S_n$ we denote the
$i$th element of $s$ by $s_i$, write sometimes $s$ as $(s_i, s_{-i})$,
and use the following standard notation:

\begin{itemize}
\item $s_{-i} := (s_1, \LL, s_{i-1}, s_{i+1}, \LL, s_n)$,

\item $S_{-i} := S_1 \times \LL \times S_{i-1} \times S_{i+1} \times \LL \times S_n$.

\end{itemize}


In the remainder of the paper we assume an initial strategic game
\[
H := (H_1, \LL, H_n, p_1, \LL, p_n).
\]
A \oldbfe{restriction} of $H$ is a sequence $(G_1, \LL, G_n)$ such that
$G_i \sse H_i$  for all $i \in \{1, \LL, n\}$. We identify the restriction 
$(H_1, \LL, H_n)$ with $H$.
We shall focus on the complete lattice
that consists of the set of all restrictions of the game $H$
ordered by the componentwise set inclusion:
\[
\mbox{$(G_1, \LL, G_n) \sse (G'_1, \LL, G'_n)$ iff $G_i \sse G'_i$ for all $i \in \{1, \LL, n\}$.}
\]
So $H$ is the largest element in this lattice and
$\bigcap_{\alpha < \gamma}$ is the customary set-theoretic 
operation on the restrictions.

Consider now a restriction $G := (G_1, \LL, G_n)$ of $H$
and two strategies $s_i, s'_i$ from $H_i$ (so \emph{not necessarily} from $G_i$). 
We say that 
$s_i$ \oldbfe{is strictly dominated} \oldbfe{on} $G$ by $s'_i$
(and write $s'_i \succ_{G} s_i$)
if
\[
\fa s_{-i} \in G_{-i} \: p_{i}(s'_i, s_{-i}) > p_{i}(s_i, s_{-i}),
\]
and that $s_i$ \oldbfe{is weakly dominated} \oldbfe{on} $G$ by $s'_i$
(and write $s'_i \succ^{w}_{G} s_i$)
if
\[
\fa s_{-i} \in G_{-i} \: p_{i}(s'_i, s_{-i}) \geq p_{i}(s_i, s_{-i}) \A \te s_{-i} \in G_{-i} \: p_{i}(s'_i, s_{-i}) > p_{i}(s_i, s_{-i}).
\]







Given a restriction $G' := (G'_1, \LL, G'_n)$ of $H$, we say that the
strategy $s_i$ from $H_i$ is a \oldbfe{best response in $G'$ to
  $s_{-i} \in G_{-i}$} if
\[
\fa s'_i \in G'_i \:
p_i(s_i, s_{-i}) \geq p_i(s'_i, s_{-i}).
\]

\subsection{Possibility correspondences}
\label{subsec:poss}

In this and the next subsection we recall the basic notions, following \cite{BB99}.
Fix a non-empty set $\Omega$ of \oldbfe{states}.
By an \oldbfe{event} we mean a subset of $\Omega$.

A \oldbfe{possibility correspondence} is a mapping from $\Omega$ to 
the powerset ${\cal P}(\Omega)$ of $\Omega$.
The following three properties of a possibility correspondence $P$ are relevant:
\begin{enumerate}\smallromani
\item for all $\omega$, $P(\omega) \neq \ES$,
\item for all $\omega$ and $\omega'$, $\omega' \in P(\omega)$ implies $P(\omega') = P(\omega)$,
\item for all $\omega$, $\omega \in P(\omega)$.
\end{enumerate}

If the possibility correspondence satisfies properties (i)--(iii), we
call it a \oldbfe{knowledge correspondence}.\footnote{In the modal
  logic terminology a knowledge correspondence is a frame for the
  modal logic S5, see, e.g.  \cite{BRV01}.}  Note that each knowledge
correspondence $P$ yields a partition $\{P(\omega) \mid \omega \in
\Omega\}$ of $\Omega$.

\subsection{Models for games}

We now link possibility correspondences with strategic games.  
Given a
restriction $G := (G_1, \LL, G_n)$ of the initial game $H$, by a
\oldbfe{model} for $G$ we mean a set of states $\Omega$ together with
a sequence of functions $\mybar{s_i}: \Omega \myra G_i$, where $i \in
\{1, \LL, n\}$. We denote it by $(\Omega, \mybar{s_1}, \LL, \mybar{s_n})$.

In what follows, given a function $f$ and a subset $E$ of
its domain, we denote by $f(E)$ the range of $f$ on $E$ and by $f
\restr{E}$ the restriction of $f$ to $E$. 

By the \oldbfe{standard model} ${\cal M}$ for $G$ we mean the model in which

\begin{itemize}
 \item $\Omega := G_1 \times \LL \times G_n$ (which means that for
   $\omega \in \Omega$, $\omega_i$ is well-defined),
 \item $\mybar{s_i}(\omega) := \omega_i$.
\end{itemize}
So the states of the standard model for $G$ are exactly the joint strategies in $G$,
and each $\mybar{s_i}$ is a projection function.
Since the initial game $H$ is given, we know the payoff functions $p_1,
\LL, p_n$. So in the context of $H$ a standard model is just an alternative way of
representing a restriction of $H$.

Given a (not necessarily standard) model ${\cal M} := (\Omega, \mybar{s_1}, \LL, \mybar{s_n})$ for a restriction
$G$ and a vector of events $\overline{E} = (E_1, \LL, E_n)$ in ${\cal
  M}$ we define
\[
G_{\overline{E}} := (\mybar{s_1}(E_1), \LL, \mybar{s_n}(E_n))
\]
and call it the \oldbfe{restriction of $G$ to $\overline{E}$}.
When each $E_i$ equals $E$ we write $G_{E}$ instead of $G_{\overline{E}}$.

Finally, we extend the notion of a model for a restriction $G$ to a
\oldbfe{knowledge model} for $G$ by assuming that each player $i$ has
a knowledge correspondence $P_i$ on $\Omega$.

\section{Local and global properties}
\label{sec:setup}


  


Given player $i$ in the initial strategic game $H := (H_1, \LL, H_n, p_1,
\LL,p_n)$ we formalize his notion of optimality using a property
$\phi_i(s_i, G)$ that holds between a strategy $s_i \in G_i$ and a
restriction $G$ of $H$.  Intuitively, $\phi_{i}(s_i, G)$
holds if $s_i$ is an `optimal' strategy for player $i$ within the
restriction $G$, assuming that he uses the property
$\phi_i$ to select optimal strategies.

We distinguish  between what we call `local' and `global' optimality.
This terminology is from \cite{Apt07}.
To assess optimality of a strategy $s_i$ \emph{locally} within the restriction $G$,
it is sufficient for $i$ to compare $s_i$ with \emph{only those strategies} $s'_i$
\emph{that occur in} $G_i$.  On the other hand, to assess the optimality of
$s_i$ \emph{globally}, player $i$ must consider all of his strategies $s'_i$ that occur
in his strategy set $H_i$ in the \emph{initial game} $H$.

Global properties are then those in which a player's strategy is evaluated
with respect to all his strategies in the initial game, whereas local properties
are concerned solely with a comparison of strategies available in the restriction $G$.
We will write $\phi^l$ when we refer to a local property, and $\phi^g$
when we refer to a global property.
Here are some natural examples.
We also give one example in both its
local and global form in order to illustrate the distinction between them:

\begin{itemize}
  
\item $sd^l_{i}(s_i, G)$ that holds iff the strategy $s_i$ of
  player $i$ is not strictly dominated on $G$ by any strategy from $G_i$
  (i.e., $\neg \te s'_i  \in G_i \: s'_i \succ_{G} s_i$),

\item $sd^g_{i}(s_i, G)$ that holds iff the strategy $s_i$ of
  player $i$ is not strictly dominated on $G$ by any strategy from $H_i$
  (i.e., $\neg \te s'_i  \in H_i \: s'_i \succ_{G} s_i$),


\item $wd^l_{i}(s_i, G)$ that holds iff the strategy $s_i$ of
  player $i$ is not weakly dominated on $G$ by any strategy from $G_i$
  (i.e., $\neg \te s'_i \in G_i \: s'_i \succ^{w}_{G} s_i$),


\item $br^l_{i}(s_i, G)$ that holds iff the strategy $s_i$ of
  player $i$ is a best response among $G_i$ to some $s_{-i} \in G_{-i}$,
i.e., 
  $\te s_{-i} \in G_{-i} \: \fa s'_i \in G_i \: p_i(s_i, s_{-i}) \geq p_i(s'_i, s_{-i})$).
\end{itemize}


Each sequence of properties $\overline{\phi} := (\phi_1, \LL, \phi_n)$
determines an operator $T_{\overline{\phi}}$ on the restrictions of
$H$ defined by
\[
T_{\overline{\phi}}(G) := (G'_1, \LL, G'_n),
\]
where $G := (G_1, \LL, G_n)$ and for all $i \in \{1, \LL, n\}$
\[
G'_i := \{ s_i \in G_i \mid \phi_i(s_i, G)\}.
\]

Since $T_{\overline{\phi}}$ is contracting, by Note \ref{note:con} it
has an outcome, i.e., $T_{\overline{\phi}}^{\infty}$ is well-defined.
$T_{\overline{\phi}}(G)$ is the result of removing from $G$ all
strategies that are not $\phi_i$-optimal. So the outcome of
$T_{\overline{\phi}}$ is the result of the iterated elimination of
strategies that for player $i$ are not $\phi_i$-optimal, where $i \in
\{1, \LL, n\}$.  In the literature, see, e.g. \cite{Lip91},
\cite{CLL07} and \cite{Apt07}, examples are given showing that for
the above optimality notions in general transfinite iterations are
necessary to reach the outcome.

When each property $\phi_i$ equals ${\textit{sd}^{\: l}}$, we write
$T_{{\textit{sd}^{\: l}}}$ instead of $T_{\overline{{\textit{sd}^{\:
        l}}}}$ and similarly with other specific properties.

%








In the next section we assume that each player $i$ employs some
property $\phi_i$ to select his strategies, and we analyze the situation
in which this information becomes commonly known via public announcements.
To determine which strategies are then selected by the players we shall use the $T_{\overline{\phi}}$ operator.  

\section{Public announcements of optimality}
\label{sec:opt}

We show that for a large class of properties $\phi_i$ the outcome
$T_{\overline{\phi}}^{\infty}$ of the iterated elimination of
strategies that for player $i$ are not $\phi_i$-optimal can be
characterized by means of the concept of a \emph{public announcement}.
This approach, inspired by \cite{vB07}, applies to all global
properties introduced in Section \ref{sec:setup}.

The particular kind of ``public announcement'' that we will be
interested in is a set of true statements, one by each player $i$ to
the effect that $i$ will not play any strategy that is not optimal for
him, according to his notion of optimality.  Note that there is no
strategic element to these announcements: the players simply follow a
protocol from which they cannot deviate.  The announcements are
``public'' in the sense that every other player 'hears' them as they
happen.

The iterated public announcements can be thought of as a process in
which the players \emph{learn} how the game will be played.  The limit
of this learning process represents the situation in which the
announcements lead to no change in the model, at which point it can be
said that \emph{rationality} has been learned by all players. It is in this
sense that public announcements provide alternative epistemic
foundations for the outcome $T_{\overline{\phi}}^{\infty}$.

Let us clarify first what we would like to achieve.  
Consider a model
${\cal M}$ for the initial game $H$. The process of iterated
elimination of the strategies that are not $\phi_i$-optimal,
formalized by the iterated applications of the $T_{\overline{\phi}}$
operator, produces a sequence $T^{\alpha}_{\overline{\phi}}$, where
${\alpha}$ is an ordinal, of restrictions of $H$. We would like to
mimic it on the side of the models, so that we get a corresponding
sequence ${\cal M}^{\alpha}$ of models of these restrictions.

To make this idea work we need to define an appropriate way of
reducing models. We take care of it by letting the players repeatedly
announce that they only select $\phi_i$-optimal strategies.  This
brings us to the notions of public announcements and their effects on
the models.

Given a model ${\cal M} = (\Omega, \mybar{s_1}, \LL, \mybar{s_n})$ for
some restriction of $H$ we define 

\begin{itemize}

\item a \oldbfe{public announcement} by player $i$ in a model ${\cal M}$
as an event $E$ in ${\cal M}$,

\item given a vector $\overline{E} := (E_1, \LL, E_n)$ of public announcements by
players $1, \LL, n$
we let
\[
[\overline{E}]({\cal M}) := (\cap_{i = 1}^{n} E_i, (\mybar{s_j} \restr{\cap_{i = 1}^n
E_i})_{j \in \{1, \LL, n\}})
\]
and call it \oldbfe{the effect of the public announcements of
  $\overline{E}$ on ${\cal M}$}.
\end{itemize}

Given a property $\phi_i$ that player $i$ uses to select his
strategies in the restriction $G$ of $H$ and a model ${\cal M} :=
(\Omega, \mybar{s_1}, \LL, \mybar{s_n})$ for $G$ we define $\newMS{\phi_i}$ as the
event in ${\cal M}$ that player $i$ selects optimally his strategies
with respect to $G$.  Formally:
\[
\newMS{\phi_i} := \{\omega \in \Omega \mid \phi_i(\mybar{s_{i}}(\omega), G)\}.
\]
We abbreviate the vector
$(\newMS{\phi_1}, \LL, \newMS{\phi_n})$ to $\newMS{\overline{\phi}}$.

We want now to obtain the reduction of a model ${\cal M}$ of $G$ to a
model ${\cal M}$ of $T_{\overline{\phi}}(G)$ by means of the just
defined vector $\newMS{\overline{\phi}}$ of public announcements.  

The effect of the public announcements of $\overline{E}$ on a model of
$G$ should ideally be a model of the restriction $G_{\overline{E}}$.
Unfortunately, this does not hold in such generality.  Indeed, let the
two-player game $G$ have the strategy sets $G_1 := \{U, D\}$, $G_2 :=
\{L, R\}$ and consider the model $\mathcal{M}$ for $G$ with $\Omega :=
\{\omega_{ul}, \omega_{dr}\}$ and the functions $\mybar{s_1}$ and $\mybar{s_2}$
defined by
\[
\mybar{s_1}(\omega_{ul}) = U, \ \mybar{s_2}(\omega_{ul}) = L, \ \mybar{s_1}(\omega_{dr}) = D, \ \mybar{s_2}(\omega_{dr}) = R.
\]
Let $\overline{E} = (\{\omega_{ul}\}, \{\omega_{dr}\})$. Then
$[\overline{E}](\cal{M}) = \emptyset$, which is not a model of $G_{\overline{E}} =
(\{U\}, \{R\})$.
This simple example is depicted in Figure \ref{fig:ceg}.

\begin{figure}[htbc]
\begin{center}
\begin{game}{2}{2}
        & $L$                & $R$                \\
$U$        & $\omega_{ul}$        &                 \\
$D$        &                 & $\omega_{dr}$
\end{game}
\caption{\label{fig:ceg}A motivating example for the use of standard models}
\end{center}
\end{figure}

A remedy lies in restricting one's attention to standard models.
However, in order to find a faithful public announcement analogue to strategy elimination
we must also
narrow the concept of a public announcement as follows.
A \oldbfe{proper public announcement} by player $i$ in a
standard model is a subset of
$\Omega = G_1 \times \LL \times G_n$ of the form $G_1 \times \LL
\times G_{i-1} \times G'_i \times G_{i+1} \times \LL \times G_n$.

So a proper public announcement by a player is an event that amounts
to a `declaration' by the player that he will limit his attention to a
subset of his strategies, that is, will discard the remaining
strategies.  So when each player makes a proper public announcement,
their combined effect on the standard model is that the set of states
(or equivalently, the set of joint strategies) becomes appropriately
restricted.  An example, which is crucial for us, of a proper public
announcement in a standard model is of course $\newMS{\phi_i}$.

The following note links in the desired way two notions we introduced.
It states that the effect of the proper public announcements of
$\overline{E}$ on the standard model for $G$ is the standard model
for the restriction of $G_{\overline{E}}$.

\begin{note} \label{not:link2}
  Let ${\cal M}$ be the standard model for $G$ and $\overline{E}$ a
  vector of proper public announcements by players $1, \LL, n$ in
  ${\cal M}$.  Then $[\overline{E}]({\cal M})$ is the standard model
  for $G_{\overline{E}}$.
\end{note}

\Proof
We only need to check that $\cap_{i = 1}^{n} E_i$ is the set of joint strategies of
the restriction $G_{\overline{E}}$.
But each $E_i$ is a proper announcement, so it is of the form 
$G_1 \times \LL \times G_{i-1} \times G'_i \times G_{i+1} \times \LL \times G_n$, 
where $G = (G_1, \LL, G_n)$. So $\cap_{i = 1}^{n} E_i = G'_1 \times \LL \times G'_n$. 

Moreover, each function $\mybar{s_i}$ is a projection, so
\[
G_{\overline{E}} = (\mybar{s_1}(E_1), \LL, \mybar{s_n}(E_n)) = (G'_1, \LL, G'_n).
\] 
\HB \VV

We also have the following observation that links the vector
$\newMS{\overline{\phi}}$ of public announcements with the operator
$T_{\overline{\phi}}$ of Section \ref{sec:setup}.

\begin{note} \label{not:link}
  Let ${\cal M} := (\Omega, \mybar{s_1}, \LL, \mybar{s_n})$ be the standard model for $G$. Then
\[
T_{\overline{\phi}}(G) = G_{[\![\overline{\phi}]\!]}.
\]
\end{note}
\Proof
Let $G = (G_1, \LL, G_n)$, $T_{\overline{\phi}}(G) = (G'_1, \LL, G'_n)$ and
$G_{[\![\overline{\phi}]\!]} = (S''_1, \LL, S''_n)$.

Fix $i \in \{1, \LL, n\}$. Then we have the following string of equivalences:
\[
\begin{array}{lll}
&              & s_i \in G'_i  \\
& \textrm{iff} & s_i \in G_i \A \phi_i(s_i, G) \\
\textrm{($\mybar{s_i}$ is onto)} & \textrm{iff} & s_i \in G_i \A \te \omega \in \Omega \: (s_i = \mybar{s_i}(\omega) \A \phi_i(\mybar{s_i}(\omega), G)) \\
& \textrm{iff} & s_i \in G_i \A \te \omega \in \newMS{\phi_i} \: (s_i = \mybar{s_i}(\omega)) \\
& \textrm{iff} & s_i \in S''_i.
\end{array}
\]
\HB
\VV

Denote now by $\newMS{\overline{\phi}}^{\infty}$ 
the iterated effect of the public
announcements of $\newMS{\overline{\phi}}$ 
starting with the standard model
for the initial game $H$.
The following conclusion then relates the iterated elimination of the
strategies that for player $i$ are not $\phi_i$-optimal to the iterated effects of
the corresponding public announcements.

\begin{theorem}
\label{thm:optanc}
$\newMS{\overline{\phi}}^{\infty}$ is the standard model for the
restriction $T_{\overline{\phi}}^{\infty}$.
\end{theorem}
\Proof
By Notes \ref{not:link2} and \ref{not:link}.
\HB
\VV

Note that in the above theorem each effect of the
public announcements of $\newMS{\overline{\phi}}$ is considered on
a different standard model.  


\section{Public announcements of rationality}
\label{sec:rat}

The above analysis gives an account of public announcements of the
\emph{optimality} of players' strategies.  We now extend this analysis to
public announcements of \emph{rationality} which takes into account
players' knowledge. To this end we
additionally assume for each player a knowledge correspondence $P_i:
\Omega \rightarrow {\cal P}(\Omega)$.

We define then the event of player $i$ being $\phi_i$-\oldbfe{rational} in the
restriction $G$ as
\[
\eventra{\phi_i} := \{\omega \in \Omega \mid \phi_i(\mybar{s_{i}}(\omega), G_{P_i(\omega)})\}.
\]

Note that when player $i$ knows that the state is in $P_i(\omega)$,
the restriction $G_{P_i(\omega)}$ represents his knowledge about the
players' strategies.  That is, $G_{P_i(\omega)}$ is the game he knows
to be relevant to his choice.  Hence $\phi_i(\mybar{s_{i}}(\omega),
G_{P_i(\omega)})$ captures the idea that if player $i$ uses $\phi_i$
to select his optimal strategy in the game he considers relevant, then
in the state $\omega$ he indeed acts `rationally'.

Again we abbreviate $(\eventra{\phi_1}, \ldots, \eventra{\phi_n})$ to
$\eventra{\overline{\phi}}$.
Note that $\eventra{\overline{\phi}}$ depends on the underlying knowledge model
$(\Omega, \mybar{s_1}, \LL, \mybar{s_n}, P_1, \LL, P_n)$ and on $G$. 

We extend the definition of the effect of the public announcements
$\overline{E} := (E_1, \LL, E_n)$ to knowledge models in the natural way,
by restricting each possibility correspondence to the intersection of
the events in $\overline{E}$:
\[
 [\overline{E}](\mathcal{M}, P_1, \LL, P_n) = ([\overline{E}]\mathcal{M}, P_1
\restr{\cap_{i = 1}^nE_i}, \LL, P_n \restr{\cap_{i = 1}^nE_i}).
\]
This definition is in the same spirit as in \cite{Pla89}
and in \cite[ page 72]{OR94}, where it is used
in the analysis of the puzzle of the hats.

We aim to find a class of knowledge models for which, under a mild
restriction on the properties $\phi_i$, $\langle
\overline{\phi}\rangle^\infty$, the iterated effect of the public
announcements of $\langle \overline{\phi}\rangle$ starting with the
standard knowledge model for the initial game $H$, will be the standard
knowledge model for $T^\infty_{\overline{\phi}}$.  We will therefore use
a natural choice of possibility correspondences, which we call the
\oldbfe{standard possibility correspondences}:
\[
 P_i(\omega) = \{\omega' \in \Omega \mid \mybar{s_i}(\omega) = \mybar{s_i}(\omega')\}.
\]
In particular, for the standard model
\[P_i(\omega) := \{\omega' \mid \omega_i = \omega'_i\}\]

By the \oldbfe{standard knowledge model} for a restriction  
$G$ we now mean the standard model for $G$ endowed with
the standard possibility correspondences.

The following observation holds.

\begin{note} \label{not:poss}
Consider the standard knowledge model $(\Omega, \mybar{s_1}, \LL, \mybar{s_n}, P_1, \LL, P_n)$
for a restriction $G := (G_1, \LL, G_n)$ of $H$ and a state
$\omega \in \Omega$. Then for all $\omega \in \Omega$
\[
G_{P_i(\omega)} = (G_1, \LL, G_{i-1}, \{\omega_i\}, G_{i+1}, \LL,  G_n).
\]
\end{note}
\Proof
Immediate by the fact that in the standard knowledge model for each
possibility correspondence we have
$P_i(\omega) = \{\omega' \in \Omega \mid \omega'_i = \omega_i\}$.
\HB
\VV

Intuitively, this observation states that in each state of a standard
knowledge model each player knows his own choice of strategy but knows
nothing about the strategies of the other players.  So standard
possibility correspondences represent the knowledge of each player after
he has privately selected his strategy but no information between the
players has been exchanged.  It is in that sense that the standard
knowledge models are natural.  In \cite{vB07} in effect only such
models are considered.

A large class of properties $\phi_i$ satisfy the following restriction:
\begin{description}
 \item[A] For all $(G_1, \ldots, G_n)$ and $G'_i$

$\phi_i(s_i, (G_i, G_{-i}))\: \lra \: \phi_i(s_i, (G'_i, G_{-i}))$.
\end{description}

That restriction on the properties 
$\phi_i$ is sufficient to obtain the following analogue of
Theorem \ref{thm:optanc} for the case of public 
announcements of \emph{rationality}.

\begin{theorem}
\label{thm:ratanc}
Suppose that each property $\phi_1, \LL, \phi_n$ satisfies \textbf{A}.
Then $\eventra{\overline{\phi}}^\infty$ is the standard knowledge
model for the restriction $T^\infty_{\overline{\phi}}$.
\end{theorem}
\Proof
Notice that it suffices to prove for each restriction $G$
the following statement for each $i$:

\begin{equation}
\forall \omega \in \Omega \: (\phi_i(\omega_i, G)\:\lra\:\phi_i(\omega_i,
G_{P_i(\omega)})).
  \label{equ:equal}
\end{equation}

Indeed, (\ref{equ:equal}) entails that $\eventra{\phi_i} =
\newMS{\phi_i}$, in which case the result follows from Theorem
\ref{thm:optanc} and the observation that the possibility
correspondences are restricted in the appropriate way.

But (\ref{equ:equal}) is a direct consequence of the assumption of \textbf{A}
and of Note \ref{not:poss}.
\HB
\VV

To see the consequences of the above result
note that \textbf{A} holds for each global property
$\textit{sd}_{i}^{\: g}$,
$\textit{wd}_{i}^{\: g}$,
and
$\textit{br}_{i}^{\: g}$ introduced in Section
\ref{sec:setup}.  

For each $\phi_i$ equal $\textit{sd}_{i}^{\: g}$ and finite games
Theorem \ref{thm:ratanc} boils down to Theorem 7 in \cite{vB07}. The
corresponding result for each $\phi_i$ equal to $\textit{br}_{i}^{\: g}$
and finite games is mentioned at the end of Section 5.4 of that paper.


It is important to note that the above 
Theorem does not hold for the
corresponding local properties $\textit{sd}_{i}^{\: l}$,
$\textit{wd}_{i}^{\: l}$,
or $\textit{br}_{i}^{\: l}$ introduced in Section \ref{sec:setup}.
Indeed, for each such property $\phi_{i}$ we have $\phi_i(\omega_i, G)$ 
for each state $\omega_i$ and restriction $G$ in which player $i$ has just 
one strategy. So by Note \ref{not:poss} 
$\phi_{i}(\omega_i, G_{P_i(\omega)})$ holds for each state $\omega$
and restriction $G$. 
Consequently $\eventra{\overline{\phi}} = \Omega$. 
So when each $\phi_i$ is a local property listed above,
$\eventra{\overline{\phi}}$ is an identity operator on the standard
knowledge models, that is
$\eventra{\overline{\phi}}^\infty$ is the standard knowledge
model for the initial game $H$ and not $T^\infty_{\overline{\phi}}$.

\bibliography{/ufs/apt/bib/e,/ufs/apt/bib/apt}
\bibliographystyle{handbk}

\end{document}